\documentclass[journal, twocolumn]{IEEEtran}
\usepackage{graphicx}
\usepackage{dblfloatfix}
\usepackage{sidecap}
\usepackage{dcolumn}
\usepackage{bm}
\usepackage[utf8]{inputenc}
\usepackage[T1]{fontenc}
\usepackage{mathptmx}
\usepackage{textcomp}
\usepackage{siunitx}
\usepackage{bigints}
\usepackage{setspace}
\usepackage{hyperref}
\usepackage[switch]{lineno}
\usepackage[table,xcdraw,dvipsnames]{xcolor}
\usepackage{booktabs}
\usepackage{float}
\usepackage{color}
\usepackage{ulem}
\usepackage{multirow}

\begin{document}

\title{YSGAG: The Ideal Substrate for YIG in Quantum Magnonics}

\author{\IEEEauthorblockN{Rostyslav~O.~Serha$^{1,2}$\IEEEauthorrefmark{1}, Carsten~Dubs$^3$, Christo Guguschev$^4$, Bernd~Aichner$^{1}$, David~Schmoll$^{1,2}$, Julien Schäfer$^5$,\\ Jaganandha Panda$^6$, Matthias Weiler$^5$, Philipp Pirro$^5$, Michal~Urbánek$^6$, and Andrii~V.~Chumak$^1$\IEEEauthorrefmark{2}}\vspace{2.5mm}

\IEEEauthorblockA{
$^1$ Faculty of Physics, University of Vienna, 1090 Vienna, Austria.
\\
$^2$ Vienna Doctoral School in Physics, University of Vienna, 1090 Vienna, Austria.
\\
$^3$ INNOVENT e.V. Technologieentwicklung, 07745 Jena, Germany.
\\
$^4$ Leibniz-Institut für Kristallzüchtung, 12489 Berlin, Germany.
\\
$^5$ Fachbereich Physik and Landesforschungszentrum OPTIMAS,\\ RPTU Kaiserslautern-Landau, 67663 Kaiserslautern, Germany.
\\
$^6$ CEITEC BUT, Brno University of Technology, 61200 Brno, Czech Republic.\\
Email: \IEEEauthorrefmark{1}rostyslav.serha@univie.ac.at, \IEEEauthorrefmark{2}andrii.chumak@univie.ac.at}}
\maketitle
\begin{abstract}
Quantum magnonics leverages the quantum properties of magnons to advance nanoscale quantum information technologies. Ferrimagnetic yttrium iron garnet (YIG), known for exceptionally long magnon lifetimes, is a cornerstone material typically grown as thin films on gadolinium gallium garnet (GGG) for lattice matching. However, paramagnetic GGG introduces detrimental damping at low temperatures due to substrate magnetization, undermining quantum applications. Here, we study magnetic damping in a 150\,nm-thick YIG film on a yttrium scandium gallium aluminum garnet (YSGAG) substrate, a newly developed diamagnetic alternative to GGG. Using ferromagnetic resonance spectroscopy down to 30\,mK, we compare YIG/YSGAG with a conventional YIG/GGG reference system. We demonstrate that the YIG/YSGAG system maintains low damping from 300\,K to 30\,mK, with $\mathbf{\alpha = 4.29\times10^{-5}}$ at room temperature, comparable to the best YIG/GGG films and bulk YIG, with no low-temperature upturn. The diamagnetic substrate eliminates the dissipation mechanisms that dominate on magnetized GGG, preserving low magnetic damping across the full temperature range. Consequently, YSGAG serves as an ideal substrate for YIG films in quantum magnonics and is paving the way for the development of spin-wave-based quantum technologies.
\end{abstract}

\section{Introduction}
Quantum computing has the potential to transform fields such as cryptography, materials science, and artificial intelligence by providing unprecedented advances in processing power and problem-solving capabilities~\cite{Nielsen2010, Arute2019}. A key challenge in solid-state quantum technology is the integration of quantum computing on nanochips, which necessitates a nanoscale data carrier~\cite{Chumak2022}. Magnons, the quanta of spin waves, emerge as promising candidates due to their broad frequency spectrum (GHz to THz), nanoscale wavelengths, and seamless compatibility with modern device architectures~\cite{Chumak2015}.

Quantum magnonics explores the behavior of magnons at the quantum level, offering low-dissipation information transfer and enabling coherent interactions with photons, phonons, and qubits~\cite{Tabuchi2015, Yuan2022, Li2020}. These properties position quantum magnonics as a strong contender for scalable hybrid quantum networks and quantum sensing applications~\cite{Lachance-Quirion2019}. However, a key challenge remains in extending magnon lifetimes at low temperatures to match those of superconducting circuits, where long coherence times are essential for quantum computing applications~\cite{Krantz2019}.

Yttrium iron garnet (YIG), a model medium in magnonics \cite{SagaYIG,Serha2025c}, is the material of choice for quantum magnonics due to its exceptionally low magnetic damping and long magnon lifetimes at room temperature (RT), both in bulk form and as thin films grown on gadolinium gallium garnet (GGG) substrates~\cite{Dubs2017, Klingler2017}. Recently it was discovered, that at millikelvin temperatures short-wavelength dipolar-exchange magnons in bulk YIG crystals can achieve lifetimes exceeding 18 µs, comparable to those of superconducting qubits~\cite{Serha2025b} and exceeding the typical magnon lifetimes of about 1\,µs by more than an order of magnitude. 

To harness the key advantages of magnons for quantum circuits, it is essential to work with propagating single magnons that are entangled either with each other or with superconducting qubits acting as a source and a detector. Thin YIG films offer a suitable medium for this purpose, and the magnon lifetime must be sufficiently long to enable the coherent transmission of quantum information over a meaningful distance~\cite{Chumak2022}. A key limitation is that YIG films grown on GGG substrates exhibit increased magnetic damping at low temperatures, which significantly reduces the magnon lifetime and limits their use in quantum technologies.

This increased damping at cryogenic temperatures is attributed to the paramagnetic GGG substrate, which becomes easily magnetized by externally applied fields, producing a stray field that penetrates the YIG layer and leads to an inhomogeneous increase in damping \cite{Serha2024, Serha2025, Schmoll2025}. Other contributions to the increased low-temperature magnon dissipation are believed to be rare earth impurities in the YIG film or substrate, and the dipolar or exchange coupling of the magnetized GGG with the YIG spin system. These interactions lead to additional damping and a significant broadening of the ferromagnetic resonance (FMR) linewidth, ultimately limiting the performance of YIG/GGG systems in quantum applications~\cite{Guo2022, Jermain2017, Michalceanu2018, Kosen2019, Schmoll2024, Knauer2023}. 

To overcome this challenge, efforts have been made to grow thin YIG films on alternative diamagnetic substrates. Promising candidates include yttrium aluminium garnet (YAG)~\cite{Sposito2013,Krysztofik2021} and yttrium scandium gallium garnet (YSGG)~\cite{Guo2023,Legrand2024}. However, YIG films grown on these substrates have not achieved the same level of structural quality or low magnetic damping as those grown on GGG by liquid phase epitaxy (LPE) technique~\cite{Dubs2017,Dubs2025, Dubs2020}, neither at room temperature nor under cryogenic conditions~\cite{Cole2023}.

Recently, a new substrate material with improved lattice matching, up to zero lattice mismatch for YIG, was recently introduced by C. Guguschev and C. Dubs \textit{et al.}~\cite{Dubs2025}: yttrium scandium gallium aluminium garnet (YSGAG). In contrast to GGG, YSGAG is diamagnetic with concomitant temperature-independent small diamagnetic susceptibility. The absence of net magnetic moments in the diamagnetic material excludes dipolar or exchange coupling of spins of YIG and YSGAG.

Here, we demonstrate that replacing GGG with YSGAG as the substrate for YIG films eliminates the detrimental impact on magnon damping at millikelvin temperatures while preserving the material's intrinsically low damping at RT. Using broadband FMR spectroscopy, we show that the linewidths in the YIG/YSGAG systems remain constant from RT down to millikelvin temperatures. We merely observe a linewidth broadening in the temperature range between 50\,K and 4\,K, which we attribute to rare earth impurities. In contrast, we show that the YIG/GGG reference systems exhibit a significant increase of damping due to paramagnetic substrate-related effects.

These findings establish YSGAG as an ideal substrate for YIG in quantum magnonics, paving the way for operating with propagating magnons in solid-state quantum systems at cryogenic temperatures.

\section{\label{Methods}Methodology}
\begin{figure*}[h!]
    \includegraphics[width=1\linewidth]{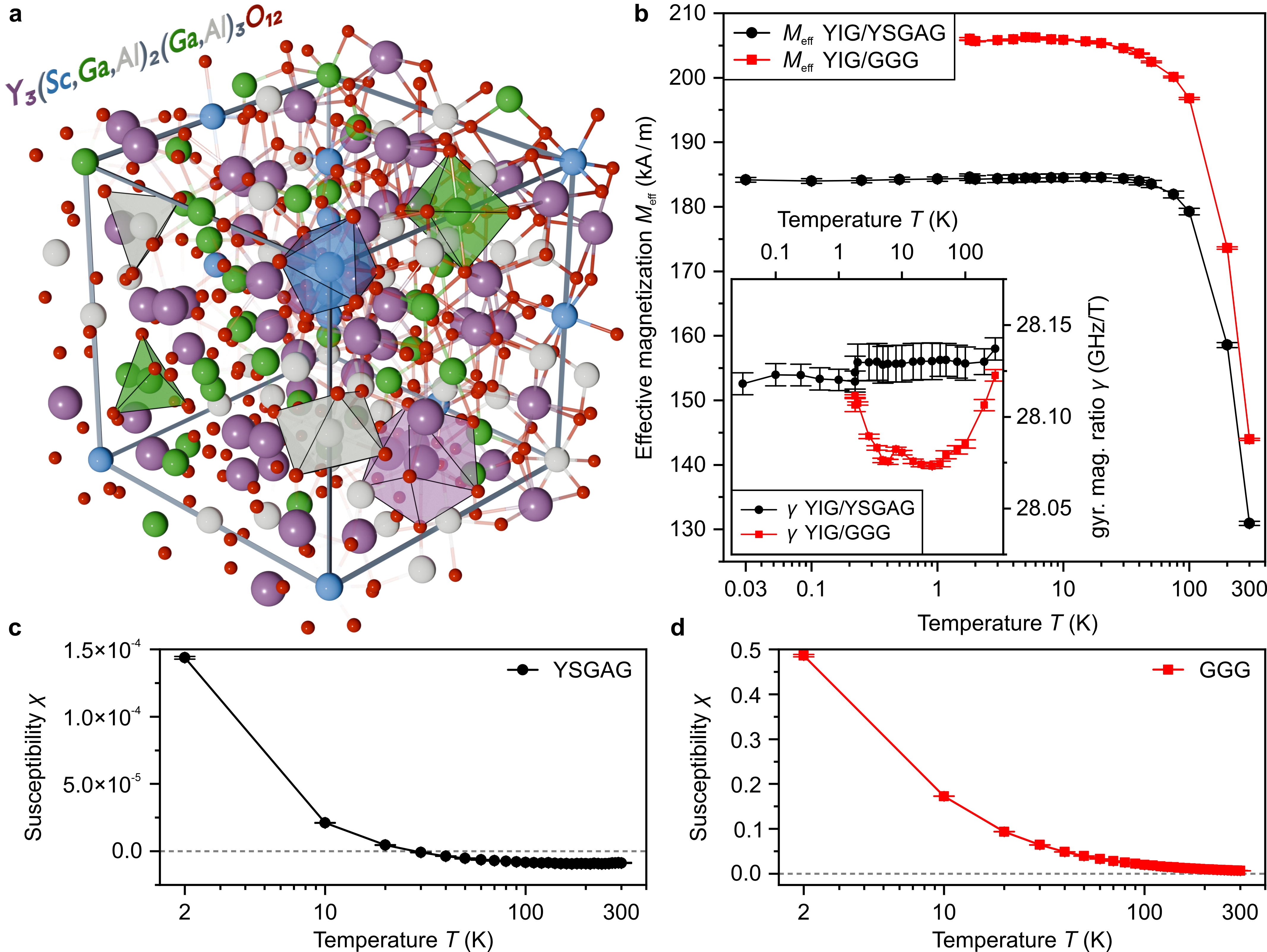}
    \caption{(a) Crystallographic structure of the diamagnetic YSGAG substrate. The cubic lattice structure of YSGAG is similar to that of YIG, but with iron ions replaced at the octahedral sites by scandium, gallium, and aluminum, while tetrahedral sites are occupied by gallium and aluminum. The samples were grown in the $\langle111\rangle$ direction, corresponding to one of the body diagonals of the cubic lattice. (b) Effective magnetization $M_{\mathrm{eff}}$ of YIG films vs temperature $T$, on an x-axis logarithmic scale for the YIG/YSGAG (black) and YIG/GGG (red) samples. For the YIG/GGG sample, the corrected $M_{\mathrm{eff}}$ could only be obtained for measurements in the PPMS down to 1.8\,K, as precise GGG magnetization data is required to correct for the GGG stray field in the Kittel fit. The inset shows the gyromagnetic ratio $\gamma$ as a function of temperature $T$, plotted on an x-axis logarithmic scale. (c) and (d) Magnetic susceptibility $\chi$ of the YSGAG substrate and GGG substrate accordingly vs the temperature $T$, plotted on an x-axis logarithmic scale. 
    }
    \label{f:1}
\end{figure*}  
In this work, we studied a 150\,nm-thick YIG ($\mathrm{Y}_3\mathrm{Fe}_5\mathrm{O}_{12}$) film grown by LPE in the $\langle111\rangle$ crystallographic direction on a 900\,µm-thick YSGAG ($\mathrm{Y}_3\mathrm{(Sc,Ga,Al)}_2\mathrm{(Ga,Al)}_3\mathrm{O}_{12}$) diamagnetic substrate, which crystallographic structure is portrayed in Fig.\,\ref{f:1}\,(a)). To compare the influence of the substrate on the YIG properties, we used a 140\,nm-thick YIG film grown by LPE on a 500\,µm-thick GGG ($\mathrm{Gd}_3\mathrm{Ga}_5\mathrm{O}_{12}$) substrate under identical conditions as a reference. The final samples were diced into $(5\times5)\,\mathrm{mm}^2$ (YIG/YSGAG) and $(7\times8)\,\mathrm{mm}^2$ (YIG/GGG) chips and characterized by broadband FMR spectroscopy using a vector network analyzer (VNA). To minimize the asymmetric influence of the stray field of paramagnetic GGG at low temperatures on the FMR linewidth \cite{Serha2025}, we microstructured 500\,µm-wide YIG stripes onto the GGG and YSGAG substrates. The exact method is detailed in~\cite{Schmoll2025} and results in a narrowing of the FMR linewidth in YIG films even at room temperature, independent of the substrate, by limiting the inhomogeneity of the YIG beneath the measurement antenna.

The measurements were performed within a Quantum Design Physical Property Measurement System (PPMS) at temperatures ranging from 2\,K to 300\,K and up to frequencies of 40\,GHz. The sample was mounted on a frequency-broadband antenna perpendicular to the YIG stripe and placed in a homogeneous magnetic field of up to 1.3\,T, generated by superconducting coils. The maximum applied microwave power in the PPMS was -5\,dBm at the VNA. For temperatures below 2\,K, measurements were conducted in a dilution refrigerator, achieving base temperatures around 10\,mK. At 20\,mK, the refrigerator provides a cooling power of 14\,µW, which is sufficient to maintain thermal equilibrium during FMR spectroscopy measurements with an applied power of -25\,dBm at the VNA.

The following measurements were performed with the external magnetic field applied along the FMR antenna. To obtain the FMR spectrum at a specific field, the transmission parameter $S_{12}$ was measured using a VNA. For both samples, measurements were performed not only at the target field, but also at two reference fields, offset by approximately 20\,mT above and below the target value~\cite{Weiler2018}. By subtracting the averaged signals at the reference fields from the measured FMR signal, we minimized the magnetically active background signal originating from the GGG substrate, obtaining the FMR absorption spectrum only from YIG~\cite{Serha2024,Serha2025}. This dual reference measurement approach enabled to obtain the best signal to noise ratio with reliable background in the measurements when working at low temperatures, since the GGG parasitic offset signal is greatly affected by the change in the applied field. To obtain the resonance frequency $f_{\mathrm{FMR}}$ and full linewidth at half maximum (FWHM) $\Delta B$, the resonance shape is fitted using a Lorentzian model. Unlike with unstructured YIG/GGG films, this approach captures the FMR lineshape across all temperatures, as microstructuring produces symmetric resonance curves~\cite{Schmoll2025}.

By linearly fitting the FWHM vs the FMR frequency we obtain the effective magnetic damping parameters, effective Gilbert damping parameter $\alpha_{\mathrm{eff}}$ and the inhomogeneous linewidth broadening $\Delta B_{0}$ (see~\cite{Boettcher2022} and the supplementary materials therein), of the magnetic material.

The Gd\textsuperscript{3+} ions in GGG have a spin of (S\,=\,7/2), resulting in a total saturation magnetization $M_{\textup{GGG}}^{\textup{s}}$ = 805\,kA/m, which is notably higher than that of YIG. As reported in the literature~\cite{Serha2025, Danilov1989} and supported by our FMR analysis, the paramagnetic GGG becomes significantly magnetized in an external magnetic field at temperatures below about 100\,K. This magnetization induces a magnetic stray field $B_{\textup{GGG}}$ in the YIG layer, which causes a shift of the YIG FMR frequencies~\cite{Danilov1989,Serha2024}. For the in-plane applied magnetic field, the FMR shift is toward lower frequencies because $B_{\textup{GGG}}$ and the applied bias field $B_{\textup{0}}$ are antiparallel. However, this is not the case for the diamagnetic YSGAG substrate. 

To accurately determine the magnetization of GGG for the Kittel fit analysis of the YIG/GGG system, we utilized vibrating-sample magnetometry (VSM) on a pure GGG slab in the temperature range from 1.8\,K to 300\,K. The raw measurement VSM data is dependent on the sample shape due to self-demagnetization, which has to be recalculated into the true material law and interpolated using the Brillouin-fit to calculate the correct field of the GGG substrate~\cite{Serha2024}.

Additionally, we used VSM to determine the magnetic susceptibility $\chi$ of the YSGAG and GGG substrates used for the YIG films in this study. The susceptibility was extracted from the slope ($\chi=\partial M/\partial H$) of the magnetization curve in the linear regime between \(50\,\mathrm{mT}\) and \(1\,\mathrm{T}\) (samples used for these measurements had the following thicknesses $t_{\mathrm{GGG}}=512\,$µm$,\ t_{\mathrm{YSGAG}}=920\,$µm).

\section{Results}
\begin{figure}[!p]
    \includegraphics[width=1\linewidth]{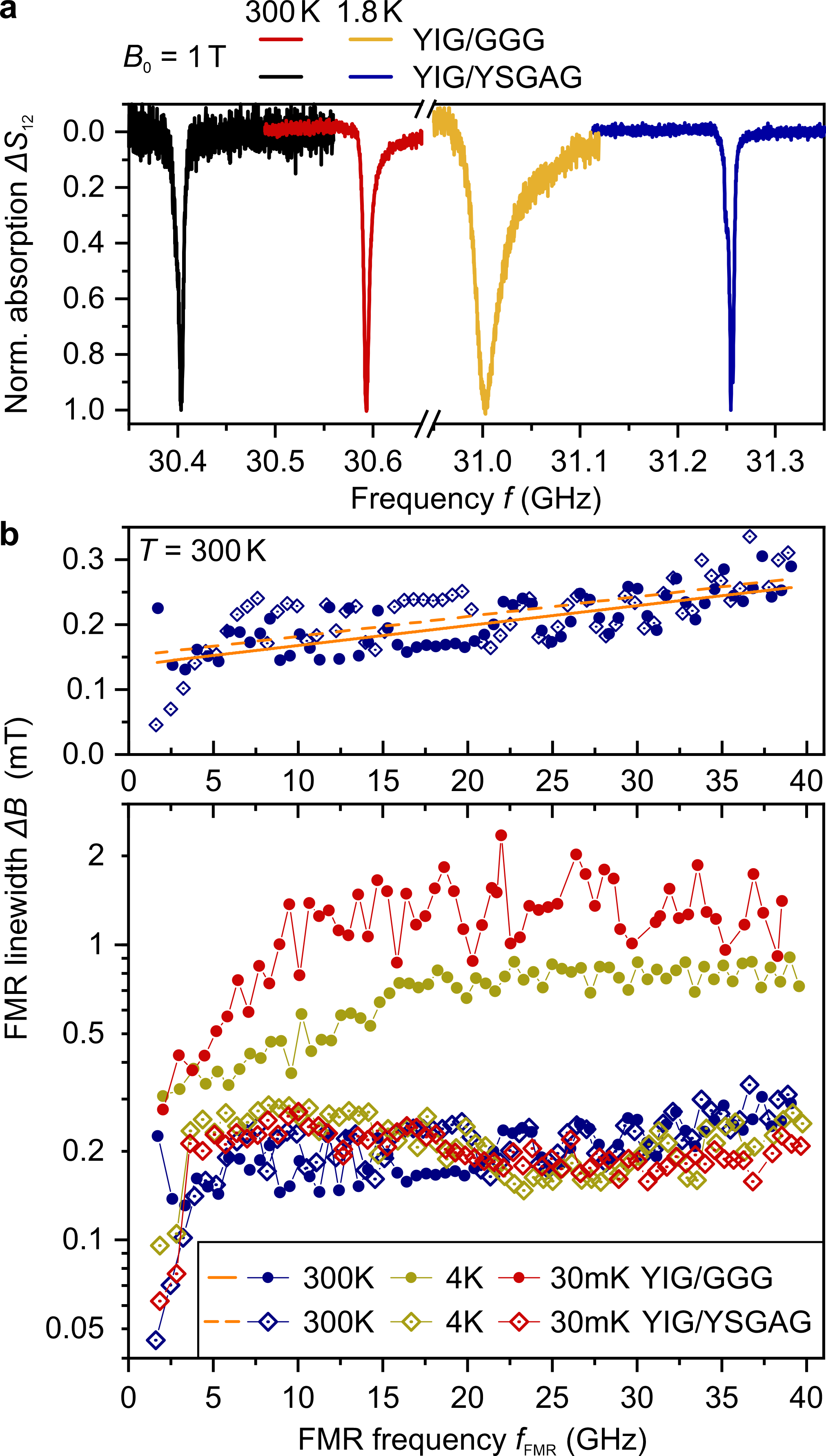}
    \caption{(a) Normalized example FMR spectra $\Delta S_{12}$ as a function of frequency $f$ for YIG/GGG and YIG/YSGAG in an in-plane oriented external field $B_0$ of 1\,T and temperatures of 300\,K and 1.8\,K. (b) (upper) FMR FWHM linewidth $\Delta B$ and corresponding Gilbert fit in orange linearly as a function of the FMR resonance frequency $f_{\mathrm{FMR}}$ at 300\,K. The closed circles and straight line represent measurements and fit for the YIG/GGG reference system, while the open diamonds and dashed line correspond to the YIG/YSGAG sample. (lower) FMR FWHM linewidth $\Delta B$ logarithmically as a function of the FMR resonance frequency $f_{\mathrm{FMR}}$ for three different temperatures. 
    }\label{f:2}
\end{figure}
Figure\,\ref{f:1}\,(a) shows the $\mathrm{Y}_3\mathrm{(Sc,Ga,Al)}_2\mathrm{(Ga,Al)}_3\mathrm{O}_{12}$ lattice of the newly developed diamagnetic substrate with selected dodecahedral, octahedral, and tetrahedral oxygen coordination polyhedra, each hosting the respective ions at their centers. A customized ratio between the $\mathrm{Y}_3\mathrm{Sc}_2\mathrm{Ga}_3\mathrm{O}_{12}$-$\mathrm{Y}_3\mathrm{Sc}_2\mathrm{Al}_3\mathrm{O}_{12}$ or $\mathrm{Y}_3\mathrm{Sc}_2\mathrm{Ga}_3\mathrm{O}_{12}$-$\mathrm{Y}_3\mathrm{Al}_5\mathrm{O}_{12}$ endmembers
allows to perfectly match the lattice parameters of the resulting YSGAG mixed crystal and the epitaxial YIG films~\cite{Dubs2025}. The results of the Kittel fit for the YIG/YSGAG sample (black) and the YIG/GGG reference sample (red), namely the effective magnetization $M_{\mathrm{eff}}$ and the gyromagnetic ratio $\gamma$ (inset), as functions of the temperature $T$ are plotted in Fig.\,\ref{f:1}\,(b). At RT, the YIG/YSGAG film exhibits a lower effective magnetization of 131\,kA/m compared to 144\,kA/m for YIG/GGG. With decreasing temperature, $M_{\mathrm{eff}}$ increases in both samples and saturates at approximately 184\,kA/m for YIG/YSGAG and 206\,kA/m for YIG/GGG. Overall, the YIG film grown on YSGAG shows about 10\,\% lower effective magnetization $M_{\mathrm{eff}}$ across the entire temperature range compared to the GGG-based reference. The YSGAG substrate used in this study has a slightly larger lattice constant ($a_{\mathrm{YSGAG}}=1.23874$\,nm) than GGG ($a_{\mathrm{GGG}}=1.2383$\,nm) \cite{Dubs2025}, thereby inducing a higher tensile strain in the YIG film compared to GGG. This, in turn, enhances the magnetostrictive anisotropy contribution, which\,—\,under in-plane tensile strain\,—\,leads to an increased uniaxial anisotropy component perpendicular to the film plane. As a result, the effective magnetization $M_{\mathrm{eff}}$ is expected to decrease. 

Below 1.8\,K, a reliable determination of $M_{\mathrm{eff}}$ for YIG/GGG was not possible, as the magnetization of the GGG substrate could not be measured in that temperature range. The gyromagnetic ratio $\gamma$ for YIG/YSGAG starts at 28.14\,GHz/T at RT and gradually decreases to 28.12\,GHz/T, consistent with previous observations in thin YIG films~\cite{Serha2024,Haidar2015,Maier-Flaig2017,Cole2023}. For YIG/GGG, the gyromagnetic ratio $\gamma$ is lower over the full temperature range, starting at 28.12\,GHz/T and decreasing to 28.07\,GHz/T at 20\,K, followed by an increase to 28.11\,GHz/T. This non-monotonic behavior at low temperatures is likely an artifact caused by the interpolation of the GGG magnetization using a Brillouin function fit. 

The susceptibility $\chi$ measured for YSGAG, presented in Fig.\,\ref{f:1}\,(c), shows that the substrate remains diamagnetic from room temperature down to 30\,K. Below this temperature it becomes weakly paramagnetic, attributed to dilute paramagnetic impurities, and reaches $\chi \approx 1.5\times10^{-4}$ at 2\,K. This value is more than three orders of magnitude smaller than the susceptibility of GGG (see Fig.\,\ref{f:1}\,(d)) and therefore has a negligible effect on the magnetic system of the YIG film, as is demonstrated by the FMR measurements presented below.

\begin{table*}[b!] 
    \caption{Important material parameters for comparison of the YIG/YSGAG system and its YIG/GGG reference.}\label{t:1}
    \includegraphics[width=1\linewidth]{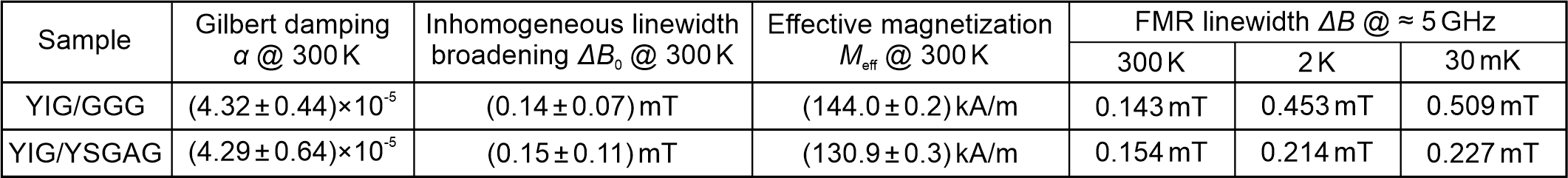}
\end{table*}

The damping behavior of the YIG/YSGAG system in comparison to YIG grown on a GGG substrate for one example external field and for the whole available frequency range is presented in Fig.\,\ref{f:2}. To demonstrate the impact of the GGG substrate on the FMR linewidth of YIG at low temperatures in comparison to the YSGAG substrate, Fig.\,\ref{f:2}\,(a) shows normalized FMR absorption spectra measured at 300\,K and 1.8\,K in an external magnetic field of $B_0 = 1\,\mathrm{T}$. At 300\,K, the FMR spectra of YIG/YSGAG and YIG/GGG are comparable, apart from the lower resonance frequency observed for the YIG/YSGAG sample, which results from a reduced effective magnetization $M_{\mathrm{eff}}$. This behavior changes significantly at lower temperatures, as illustrated by the measurements at 1.8\,K. For both systems, the FMR shifts to higher frequencies due to an increase in $M_{\mathrm{eff}}$ with decreasing temperature (see Fig.\,\ref{f:1}\,(b)). However, the resonance frequency of YIG/GGG becomes even lower than that of YIG/YSGAG, which is attributed to the stray field of the partially magnetized GGG substrate that reduces the internal field in the YIG film \cite{Danilov1989, Serha2024}. In addition, the FMR linewidth of YIG/GGG broadens significantly at low temperatures, whereas the linewidth of the YIG/YSGAG system remains visually unchanged. 

In Fig.\,\ref{f:2}\,(b), the full linewidth at half maximum of the FMR $\Delta B$ is plotted as a function of the FMR resonance frequency $f_{\mathrm{FMR}}$. The closed circles represent measurements for the YIG/GGG reference sample, while the open diamonds correspond to data from the YIG/YSGAG sample, each measured at three different temperatures: 300\,K, 4\,K, and 30\,mK. At 300\,K, both YIG/YSGAG and YIG/GGG exhibit comparable linewidths, displaying a linear dependence on frequency, which is consistent with the Gilbert damping model as shown in Fig\,\ref{f:2}\,(b upper). The extracted damping parameters are $\alpha = 4.32\times10^{-5}$ for YIG on GGG and $\alpha = 4.29\times10^{-5}$ for YIG on YSGAG, as shown in further detail and error margins in the material parameter Table\,1. The remarkably low damping at RT is further confirmed by the narrowest measured linewidth of 46\,µT at the FMR frequency of $f_{\mathrm{FMR}} = 1.6\,\mathrm{GHz}$, which corresponds to a magnon lifetime $\tau = 1/(\pi\cdot\gamma\cdot\Delta B)=$~0.25\,µs. Besides the results of the Gilbert and Kittel fits for RT, Table\,1 additionally presents the FMR linewidth $\Delta B$ at three selected temperatures for a resonance frequency $f_{\mathrm{FMR}} \approx 5\,\mathrm{GHz}$, corresponding to the frequency regime currently most relevant in quantum magnonics research.

However, as the temperature decreases, the FMR linewidth increases for YIG grown on GGG, reaching values above 2\,mT at high frequencies, while for the YIG/YSGAG $\Delta B$ remains nearly constant and even shows a slight decrease toward millikelvin temperatures. In both cases, the linewidth behavior cannot be accurately described by the Gilbert damping model throughout the whole frequency range available (see Fig\,\ref{f:2}\,(b lower)), indicating the presence of additional broadening mechanisms. For the YIG/GGG reference system, this behavior has been discussed in detail in~\cite{Serha2025,Schmoll2025}. The slight linewidth broadening observed at lower frequencies (5--18\,GHz) for the YIG/YSGAG system is attributed to the presence of double or multiple FMR peaks, which gradually differ with decreasing temperature and thereby increase the fitted linewidth. At FMR frequencies, for which these peaks can be resolved and fitted separately (up to 4\,GHz), we obtain record-low linewidths for YIG films. In this study, we analyzed the most intense FMR peaks, with linewidths below 0.1\,mT, particularly at the lowest frequencies. This contribution to linewidth broadening could be mitigated in future samples by improving material homogeneity as described in~\cite{Dubs2025,Dubs2020} and suppressing the formation of double or multi-peak FMR spectra.

\begin{figure}[!t]
    \includegraphics[width=1\linewidth]{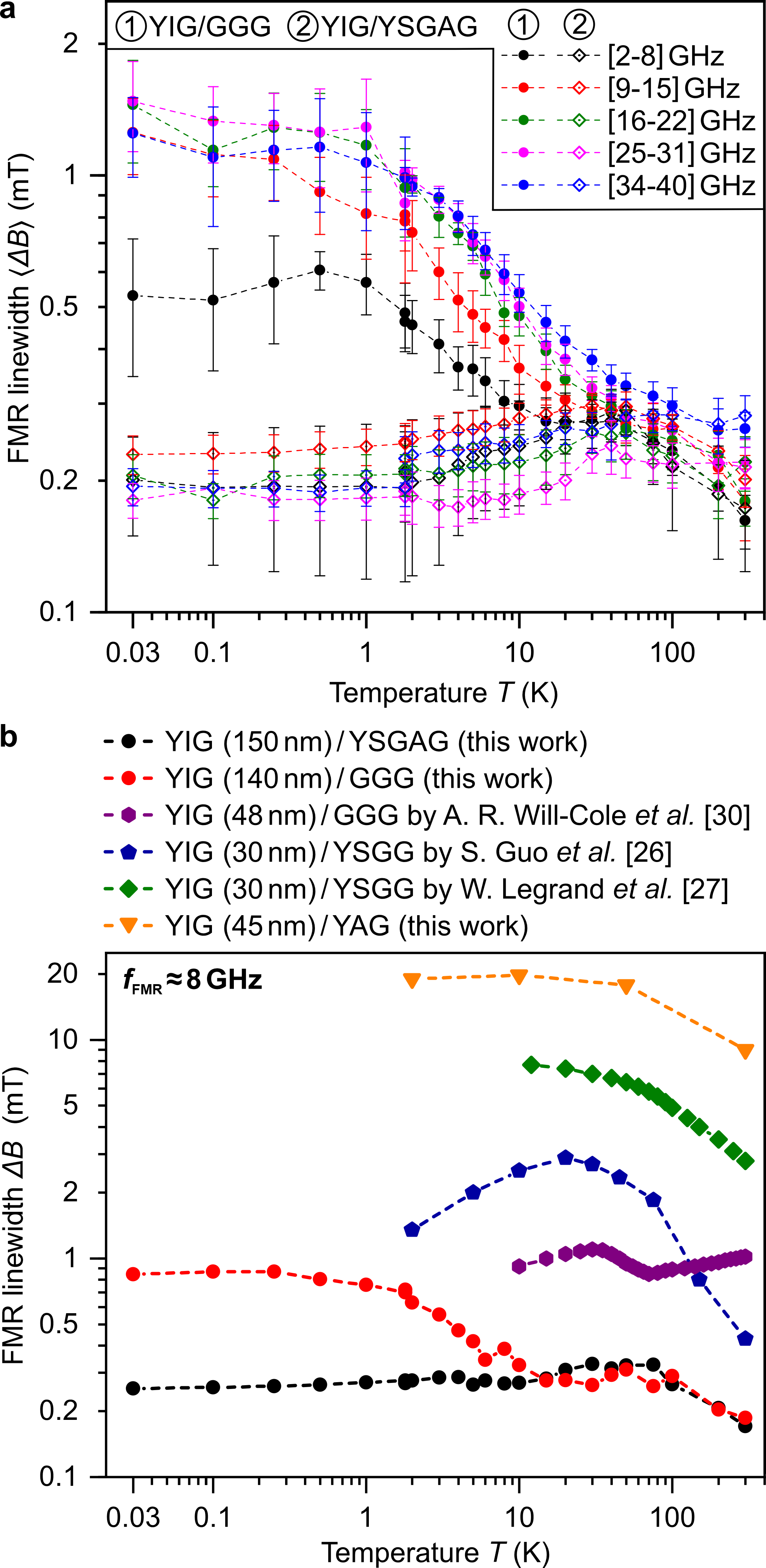}
\caption{(a) Frequency-band-averaged FMR FWHM linewidth $\langle\Delta B\rangle$ vs temperature $T$, plotted on double logarithmic scale for five different frequency bands. The closed circles represent measurements for the YIG/GGG reference system, while the open diamonds correspond to the YIG/YSGAG sample. (b) FMR FWHM linewidth $\Delta B$ as a function of temperature $T$ for an FMR frequency $f_{\mathrm{FMR}} \approx 8\,\mathrm{GHz}$, displayed on a double logarithmic scale. The graph compares measurements obtained in this work with selected results of sputtered and LPE grown YIG films from relevant literature, enabling a direct evaluation of temperature-dependent damping behavior across different samples, their substrates and studies \cite{Cole2023,Guo2023,Legrand2024}.}\label{f:3}
\end{figure}

Figure\,\ref{f:3}\,(a) shows the frequency-band-averaged FMR linewidth $\langle\Delta B\rangle$ as a function of temperature $T$, plotted on a logarithmic x-axis for five different frequency bands. Here, we average over frequency bands to reveal a clear trend in the linewidth behavior. This approach is necessary, as significant scattering in the linewidth data, particularly for YIG/GGG at low temperatures, would otherwise obscure the underlying temperature dependence (see Fig.\,\ref{f:2}\,(b, lower)). The closed circles represent measurements for the YIG/GGG reference system, while the open diamonds correspond to the YIG/YSGAG sample. With decreasing temperature, the averaged linewidth increases significantly across all frequency bands for the reference YIG/GGG sample, reaching a maximum at millikelvin temperatures. The smallest increase occurs at the lowest frequencies, where the required magnetic field is minimal and therefore magnetizes the GGG substrate less. In contrast, the averaged linewidths of YIG/YSGAG show only a slight increase, peaking around 30--50\,K and subsequently decreasing toward lower temperatures.  

As the measurements were performed using two different setups—the PPMS and the dilution refrigerator—it is worth noting that the results in the overlapping temperature range agree within each other's error bars, confirming the consistency of the data. The error bars represent the standard deviation of the mean, calculated from measurements taken over the frequency intervals in Fig.\,\ref{f:3}\,(a). Measurements conducted in the dilution refrigerator exhibited a lower signal-to-noise ratio, resulting in larger noise in the data shown in Fig.\,\ref{f:2}\,(b lower) and larger error bars in Fig.\,\ref{f:3}\,(a). To ensure the system remained in thermal equilibrium, the applied microwave power was kept below \SI{-25}{dBm}, which led to increased noise levels in the signal (see Sec.\,\ref{Methods}).  

Several mechanisms reported in the literature can be associated with the observed linewidth broadening in YIG film systems. The first and well-known mechanism, studied in both bulk and thin-film YIG, involves relaxation processes due to rare-earth impurities in GGG and YIG~\cite{Michalceanu2018, Schmoll2024, Maier-Flaig2017, Dillon1959, Spencer1959, Sparks1961, Seiden1964, Boventer2018}, and likely corresponds to the linewidth maximum observed in the YIG/YSGAG system. In the YIG/GGG reference sample, an additional broadening mechanism is the dipolar coupling between the YIG film and the partially magnetized GGG substrate, which has a large electron paramagnetic resonance (EPR) linewidth of approximately 400\,mT~\cite{Barak1992, Bedyukh1999}. However, as recently shown in~\cite{Schmoll2024}, this effect is pronounced only for propagating magnons with non-zero wavenumbers ($k \neq 0$) and is expected to vanish for FMR modes and short-wavelength exchange magnons.
For a broader perspective on the field, we compare our measurements from LPE-grown YIG films with data of sputtered YIG films on diamagnetic substrates and LPE-grown reference films on GGG substrates from the literature, specifically those by Will-Cole \textit{et al.}~\cite{Cole2023}, Guo \textit{et al.}~\cite{Guo2023}, and Legrand \textit{et al.}~\cite{Legrand2024}. This comparison is presented in Fig.\,\ref{f:3}\,(b) for a FMR frequency $f_{\mathrm{FMR}}$ of 8\,GHz. To include a YIG/YAG system in our comparison, we performed FMR measurements at four temperatures on a 45\,nm-thick sputtered YIG film on a YAG substrate (not microstructured into a stripe). The corresponding linewidths, shown in orange in Fig.\,\ref{f:3}\,(b), are the broadest in the set. At room temperature, our YIG/YAG FMR linewidth is typically $\sim\,10$\,mT, consistent with prior reports at comparable frequencies~\cite{Sposito2013,Krysztofik2021}. The worse damping parameter arises from the large lattice mismatch between YIG and YAG, which increases upon cooling and broadens the linewidth by approximately a factor of two to 20\,mT.

The observed differences in room-temperature FMR linewidths between our samples and those reported in the literature can primarily be attributed to the increased material homogeneity under the measurement antenna achieved by microstructuring the YIG films into a stripe, and to the significantly thinner YIG film thickness in those studies, by a factor of three to five, which can lead to enhanced magnetic damping. It is also worth noting that even very high quality sputtered YIG films \cite{Schmidt2020, Ding2020, Chumak2022} tend to exhibit broader FMR linewidths $\Delta B$ and/or a higher Gilbert damping $\alpha$ than LPE grown YIG films \cite{Dubs2017, Dubs2020} with high microstructural perfection.

While all of the here presented state-of-the-art YIG films demonstrate low to moderate damping at room temperature, they exhibit a marked increase in linewidth upon cooling to cryogenic temperatures. In stark contrast, the YIG/YSGAG system investigated in this work retains exceptionally narrow linewidths down to millikelvin temperatures, with only minimal temperature dependence, thereby outperforming YIG grown on GGG or YSGG substrates in the low-temperature regime.

\section{Conclusion}
Our results demonstrate that YIG films grown via LPE on diamagnetic YSGAG substrates can be compared to state of the art YIG films at RT and exhibit significantly reduced temperature-dependent magnetic damping compared to conventional YIG/GGG or YIG/YSGG systems. While the FMR linewidths of YIG/GGG increase notably at low temperatures\,—\,reaching up to 2\,mT at millikelvin temperatures\,—\,the YIG/YSGAG films maintain low damping across the entire temperature range, with linewidths remaining around 0.2\,mT. This order-of-magnitude improvement is attributed to the elimination of parasitic damping mechanisms inherent to GGG, including substrate magnetization and dipolar coupling. With further optimization of growth conditions and improved material homogeneity, the multi-peak FMR features observed in YIG/YSGAG can be minimized, allowing FMR magnon lifetimes close to those of bulk YIG to be retained even at millikelvin temperatures, as at 30\,mK we observe a linewidth $\Delta B$ of 62\,µT for an FMR frequency $f_{\mathrm{FMR}} = 1.9\,\mathrm{GHz}$.

Our study also shows that the static magnetic properties, characterized by the effective magnetization of the 150\,nm-thick YIG film on YSGAG, remain comparable to those of the reference YIG/GGG system, even at cryogenic temperatures. A consistent ~10\,\% offset of $M_\mathrm{eff}$ across the temperature range is observed, likely due to a slightly larger lattice mismatch, which offers an opportunity for further improvement through optimized lattice alignment in future samples. These results confirm that YIG/YSGAG films, with a $M_\mathrm{eff}$ of (130\,--\,185)\,kA/m (depending on the temperature) can operate within the same frequency range and technological contexts as conventional YIG/GGG systems.

In summary, the YIG/YSGAG system presents a highly promising medium for low-dissipation magnonics, enabling reliable magnetic performance down to millikelvin temperatures. These findings provide a solid foundation for accurate modeling and practical realization of next-generation magnonic devices and networks based on this magnetic system.
\section*{acknowledgements}
This research was funded in part by the Austrian Science Fund (FWF) project Paramagnonics [10.55776/I6568].
The work was supported by the German Federal Ministry of Research, Technology and Space (BMFTR) under the reference numbers (13N17108, 13N17109 and 13N17110) within a collaborative project “Low-loss materials for integrated magnonic-superconducting quantum technologies (MagSQuant)”.
MU acknowledges the support of the Grant Agency of the Czech Republic, project no. 23-04120L. Microstructuring of the sample was done at CzechNanoLab Research Infrastructure supported by MEYS CR (LM2023051).
BA acknowledges support by COST Action Polytopo CA23134 (European Cooperation in Science and 
Technology)
CD thanks Oleksii Surzhenko for room temperature magnetic measurements and R. Meyer (INNOVENT e.V.) for technical support. ROS, BA, DS, and AVC thank Adam Krysztofik for fabricating and providing the YIG/YAG sample used in this work.

\section*{Author contributions}
ROS conducted the FMR measurements on YIG/YSGAG and YIG/GGG, processed and analyzed the data, and authored the initial draft of the manuscript.
CD initiated the development of substrate crystal growth, planned the epitaxial film growth and synthesized the YIG films. 
CG developed the YSGAG substrate crystals.
BA maintained the experimental setup, supported the measurements and data analysis.  
DS maintained the experimental setup, supported the experimental measurements and conducted FMR measurements on YIG/YAG.
JS carried out measurements and further evaluation to determine the magnetic susceptibilities.
JP microstructured the YIG films.
MW, PP and MU supported the data interpretation.
AVC~planned the experiment and led the project.
All authors discussed results and contributed to the manuscript.

\section*{Competing interests}
The authors declare no competing interests.


\textbf{Data availability}
The data that support the findings of this study are available
from the corresponding authors upon reasonable request.
\emergencystretch=5em
\IEEEtriggeratref{42} 
\bibliographystyle{BST}
\bibliography{BIB}

@misc{Dubs2025,
  doi = {10.48550/ARXIV.2508.18101},
  url = {https://arxiv.org/abs/2508.18101},
  author = {Guguschev,  C. and Dubs,  C. and Blukis,  R. and Surzhenko,  O. and Br\"{u}tzam,  M. and Koc,  R. and Rhode,  C. and Berger,  K. and Richter,  C. and Berryman,  C. and Serha,  R. O. and Chumak,  A. V.},
  title = {Novel diamagnetic garnet-type substrate single crystals for ultralow-damping yttrium iron garnet $\mathrm{Y}_3\mathrm{Fe}_5\mathrm{O}_{12}$ films at cryogenic temperatures},
  publisher = {arXiv},
  year = {2025},
  copyright = {Creative Commons Attribution Non Commercial No Derivatives 4.0 International}
}

@misc{Serha2025b,
  doi = {10.48550/ARXIV.2505.22773},
  url = {https://arxiv.org/abs/2505.22773},
  author = {Serha,  Rostyslav O. and McAllister,  Kaitlin H. and Majcen,  Fabian and Knauer,  Sebastian and Reimann,  Timmy and Dubs,  Carsten and Melkov,  Gennadii A. and Serga,  Alexander A. and Tyberkevych,  Vasyl S. and Chumak,  Andrii V. and Bozhko,  Dmytro A.},
  title = {Ultra-long-living magnons in the quantum limit},
  publisher = {arXiv},
  year = {2025},
  copyright = {Creative Commons Attribution 4.0 International}
}

@article{Guo2023,
  title = {Strong on-Chip Microwave Photon–Magnon Coupling Using Ultralow-Damping Epitaxial $\mathrm{Y}_3\mathrm{Fe}_5\mathrm{O}_{12}$ Films at $2\,\mathrm{K}$},
  volume = {23},
  ISSN = {1530-6992},
  url = {http://dx.doi.org/10.1021/acs.nanolett.3c00959},
  DOI = {10.1021/acs.nanolett.3c00959},
  number = {11},
  journal = {Nano Letters},
  publisher = {American Chemical Society (ACS)},
  author = {Guo,  Side and Russell,  Daniel and Lanier,  Joseph and Da,  Haotian and Hammel,  P. Chris and Yang,  Fengyuan},
  year = {2023},
  month = may,
  pages = {5055–5060}
}

@article{Serha2025c,
  title={"{T}heory of Magnetization Temperature Dependence in Ferrimagnetics"},
  author={Serha, Rostyslav O and Pomyalov, Anna and Chumak, Andrii V and L'vov, Victor S},
  journal={arXiv preprint:2507.18209},
  year={2025},
  url = {https://arxiv.org/abs/2507.18209}
}

@article{Legrand2024,
  title = {Lattice‐Tunable Substituted Iron Garnets for Low‐Temperature Magnonics},
  ISSN = {1616-3028},
  url = {http://dx.doi.org/10.1002/adfm.202503644},
  DOI = {10.1002/adfm.202503644},
  journal = {Advanced Functional Materials},
  publisher = {Wiley},
  pages ={2503644},
  author = {Legrand,  William and Kemna,  Yana and Sch\"{a}ren,  Stefan and Wang,  Hanchen and Petrosyan,  Davit and Holder,  Luise and Schlitz,  Richard and Aguirre,  Myriam H. and Lammel,  Michaela and Gambardella,  Pietro},
  year = {2025},
  month = jun 
}

@article{Schmoll2025,
  title = {Elimination of substrate-induced ferromagnetic resonance linewidth broadening in the epitaxial system {YIG}-{GGG} by microstructuring},
  volume = {51},
  ISSN = {1090-6517},
  url = {http://dx.doi.org/10.1063/10.0036749},
  DOI = {10.1063/10.0036749},
  number = {6},
  journal = {Low Temperature Physics},
  publisher = {AIP Publishing},
  author = {Schmoll,  David and Serha,  Rostyslav O. and Panda,  Jaganandha and Voronov,  Andrey A. and Dubs,  Carsten and Urbánek,  Michal and Chumak,  Andrii V.},
  year = {2025},
  month = jun,
  pages = {724–730}
}

@article{Serha2025,
  title = {Damping enhancement in {YIG} at millikelvin temperatures due to {GGG} substrate},
  volume = {5},
  ISSN = {2950-2578},
  url = {http://dx.doi.org/10.1016/j.mtquan.2025.100025},
  DOI = {10.1016/j.mtquan.2025.100025},
  journal = {Materials Today Quantum},
  publisher = {Elsevier BV},
  author = {Serha,  Rostyslav O. and Voronov,  Andrey A. and Schmoll,  David and Klingbeil,  Rebecca and Knauer,  Sebastian and Koraltan,  Sabri and Pribytova,  Ekaterina and Lindner,  Morris and Reimann,  Timmy and Dubs,  Carsten and Abert,  Claas and Verba,  Roman and Urbánek,  Michal and Suess,  Dieter and Chumak,  Andrii V.},
  year = {2025},
  month = mar,
  pages = {100025}
}

@article{Ding2020,
  title = {Sputtering Growth of Low-Damping Yttrium-Iron-Garnet Thin Films},
  volume = {11},
  ISSN = {1949-3088},
  DOI = {10.1109/lmag.2020.2989687},
  journal = {IEEE Magn. Lett.},
  publisher = {Institute of Electrical and Electronics Engineers (IEEE)},
  author = {Ding,  Jinjun and Liu,  Tao and Chang,  Houchen and Wu,  Mingzhong},
  year = {2020},
  pages = {1–5}
}

@article{Schmoll2024,
  title = {Wavenumber-dependent magnetic losses in yttrium iron garnet--gadolinium gallium garnet heterostructures at millikelvin temperatures},
  author = {Schmoll, David and Voronov, Andrey A. and Serha, Rostyslav O. and Slobodianiuk, Denys and Levchenko, Khrystyna O. and Abert, Claas and Knauer, Sebastian and Suess, Dieter and Verba, Roman and Chumak, Andrii V.},
  journal = {Phys. Rev. B},
  volume = {111},
  issue = {13},
  pages = {134428},
  numpages = {11},
  year = {2025},
  month = {Apr},
  publisher = {American Physical Society},
  doi = {10.1103/PhysRevB.111.134428},
  url = {https://link.aps.org/doi/10.1103/PhysRevB.111.134428}
}

@article{Spencer1959,
  title = {Low-Temperature Line-Width Maximum in Yttrium Iron Garnet},
  volume = {3},
  ISSN = {0031-9007},
  url = {http://dx.doi.org/10.1103/PhysRevLett.3.32},
  DOI = {10.1103/physrevlett.3.32},
  number = {1},
  journal = {Physical Review Letters},
  publisher = {American Physical Society (APS)},
  author = {Spencer,  E. G. and LeCraw,  R. C. and Clogston,  A. M.},
  year = {1959},
  month = jul,
  pages = {32–33}
}

@article{Dillon1959,
  title = {Effects of Rare Earth Impurities on Ferrimagnetic Resonance in Yttrium Iron Garnet},
  volume = {3},
  ISSN = {0031-9007},
  url = {http://dx.doi.org/10.1103/PhysRevLett.3.30},
  DOI = {10.1103/physrevlett.3.30},
  number = {1},
  journal = {Phys. Rev. Lett.},
  publisher = {American Physical Society (APS)},
  author = {Dillon,  J. F. and Nielsen,  J. W.},
  year = {1959},
  month = jul,
  pages = {30–31}
}

@article{Sparks1961,
  title = {Ferromagnetic Relaxation. I. Theory of the Relaxation of the Uniform Precession and the Degenerate Spectrum in Insulators at Low Temperatures},
  volume = {122},
  ISSN = {0031-899X},
  url = {http://dx.doi.org/10.1103/PhysRev.122.791},
  DOI = {10.1103/physrev.122.791},
  number = {3},
  journal = {Phys. Rev.},
  publisher = {American Physical Society (APS)},
  author = {Sparks,  M. and Loudon,  R. and Kittel,  C.},
  year = {1961},
  month = may,
  pages = {791–803}
}

@article{Seiden1964,
  title = {Ferrimagnetic Resonance Relaxation in Rare-Earth Iron Garnets},
  volume = {133},
  ISSN = {0031-899X},
  url = {http://dx.doi.org/10.1103/PhysRev.133.A728},
  DOI = {10.1103/physrev.133.a728},
  number = {3A},
  journal = {Phys. Rev.},
  publisher = {American Physical Society (APS)},
  author = {Seiden,  P. E.},
  year = {1964},
  month = feb,
  pages = {A728–A736}
}

@article{Boventer2018,
  title = {Complex temperature dependence of coupling and dissipation of cavity magnon polaritons from millikelvin to room temperature},
  volume = {97},
  ISSN = {2469-9969},
  url = {http://dx.doi.org/10.1103/PhysRevB.97.184420},
  DOI = {10.1103/physrevb.97.184420},
  number = {18},
  journal = {Phys. Rev. B},
  publisher = {American Physical Society (APS)},
  author = {Boventer,  Isabella and Pfirrmann,  Marco and Krause,  Julius and Sch\"{o}n,  Yannick and Kl\"{a}ui,  Mathias and Weides,  Martin},
  year = {2018},
  month = may 
}

@article{Krysztofik2021,
  title = {Effect of strain-induced anisotropy on magnetization dynamics in $\mathrm{Y}_3\mathrm{Fe}_5\mathrm{O}_{12}$ films recrystallized on a lattice-mismatched substrate},
  volume = {11},
  ISSN = {2045-2322},
  url = {http://dx.doi.org/10.1038/s41598-021-93308-3},
  DOI = {10.1038/s41598-021-93308-3},
  number = {1},
  journal = {Sci. Rep.},
  publisher = {Springer Science and Business Media LLC},
  author = {Krysztofik,  Adam and \"{O}zoğlu,  Sevgi and McMichael,  Robert D. and Coy,  Emerson},
  year = {2021},
  month = jul 
}

@article{Sposito2013,
author = {A. Sposito and T. C. May-Smith and G. B. G. Stenning and P. A. J. de Groot and R. W. Eason},
journal = {Opt. Mater. Express},
number = {5},
pages = {624--632},
publisher = {Optica Publishing Group},
title = {Pulsed laser deposition of high-quality $\mu$m-thick {YIG} films on {YAG}},
volume = {3},
month = {May},
year = {2013},
url = {https://opg.optica.org/ome/abstract.cfm?URI=ome-3-5-624},
doi = {10.1364/OME.3.000624},
}

@article{Schmidt2020,
  author = {Schmidt, G. and Hauser, C. and Trempler, P. and Paleschke, M. and Papaioannou, E. T.},
  title = {Ultra thin films of yttrium iron garnet with very low damping: A review},
  journal = {Phys. Status Solidi (b)},
  volume = {257},
  pages = {1900644},
  year = {2020}
}

@article{Krantz2019,
  author = {Krantz, Philip and Kjaergaard, Morten and Yan, Fei and Orlando, Terry P. and Gustavsson, Simon and Oliver, William D.},
  title = {A Quantum Engineer's Guide to Superconducting Qubits},
  journal = {Appl. Phys. Rev.},
  volume = {6},
  number = {2},
  pages = {021318},
  year = {2019},
  doi = {10.1063/1.5089550}
}

@book{Nielsen2010,
  author = {Nielsen, Michael A. and Chuang, Isaac L.},
  title = {Quantum Computation and Quantum Information},
  year = {2010},
  publisher = {Cambridge~University Press},
  address = {Cambridge, UK},
  edition = {10th anniversary},
  isbn = {978-1-107-00217-3}
}

@article{Arute2019,
  author = {Arute, F. and Arya, K. and Babbush, R. and Bacon, D. and Bardin, J. C. and Barends, R. and Biswas, R. and Boixo, S. and Brandao, F. G. S. L. and Buell, D. A. and Burkett, B. and Chen, Y. and Chen, Z. and Chiaro, B. and Collins, R. and Courtney, W. and Dunsworth, A. and Farhi, E. and Foxen, B. and Fowler, A. and Gidney, C. and Giustina, M. and Graff, R. and Guerin, K. and Habegger, S. and Harrigan, M. P. and Hartmann, M. J. and Ho, A. and Hoffmann, M. and Huang, T. and Humble, T. S. and Isakov, S. V. and Jeffrey, E. and Jiang, Z. and Kafri, D. and Kechedzhi, K. and Kelly, J. and Klimov, P. V. and Knysh, S. and Korotkov, A. and Kostritsa, F. and Landhuis, D. and Lindmark, M. and Lucero, E. and Lyakh, D. and Mandr{\`a}, S. and McClean, J. R. and McEwen, M. and Megrant, A. and Mi, X. and Michielsen, K. and Mohseni, M. and Mutus, J. and Naaman, O. and Neeley, M. and Neill, C. and Niu, M. Y. and Ostby, E. and Petukhov, A. and Platt, J. C. and Quintana, C. and Rieffel, E. G. and Roushan, P. and Rubin, N. C. and Sank, D. and Satzinger, K. J. and Smelyanskiy, V. and Sung, K. J. and Trevithick, M. D. and Vainsencher, A. and Villalonga, B. and White, T. and Yao, Z. J. and Yeh, P. and Zalcman, A. and Neven, H. and Martinis, J. M.},
  title = {Quantum supremacy using a programmable superconducting processor},
  journal = {Nature},
  volume = {574},
  number = {7779},
  pages = {505-510},
  year = {2019},
  doi = {10.1038/s41586-019-1666-5}
}

@article{Tabuchi2015,
  author = {Tabuchi, Y. and Ishino, S. and Noguchi, A. and Ishikawa, T. and Yamazaki, R. and Usami, K. and Nakamura, Y.},
  title = {Coherent coupling between a ferromagnetic magnon and a superconducting qubit},
  journal = {Science},
  volume = {349},
  number = {6246},
  pages = {405-408},
  year = {2015},
  doi = {10.1126/science.aaa3693}
}

@article{Bedyukh1999,
  title = {Resonant magnetic properties of gadolinium-gallium garnet single crystals},
  volume = {25},
  ISSN = {1090-6517},
  url = {http://dx.doi.org/10.1063/1.593724},
  DOI = {10.1063/1.593724},
  number = {3},
  journal = {Low Temp. Phys.},
  publisher = {AIP Publishing},
  author = {Bedyukh,  A. R. and Danilov,  V. V. and Nechiporuk,  A. Yu. and Romanyuk,  V. F.},
  year = {1999},
  month = mar,
  pages = {182–183}
}

@article{Serha2024,
  title = {Magnetic anisotropy and GGG substrate stray field in {YIG} films down to millikelvin temperatures},
  volume = {2},
  ISSN = {2948-2119},
  url = {http://dx.doi.org/10.1038/s44306-024-00030-7},
  DOI = {10.1038/s44306-024-00030-7},
  number = {1},
  journal = {npj Spintronics},
  publisher = {Springer Science and Business Media LLC},
  author = {Serha,  Rostyslav O. and Voronov,  Andrey A. and Schmoll,  David and Verba,  Roman and Levchenko,  Khrystyna O. and Koraltan,  Sabri and Davídková,  Kristýna and Budinská,  Barbora and Wang,  Qi and Dobrovolskiy,  Oleksandr V. and Urbánek,  Michal and Lindner,  Morris and Reimann,  Timmy and Dubs,  Carsten and Gonzalez-Ballestero,  Carlos and Abert,  Claas and Suess,  Dieter and Bozhko,  Dmytro A. and Knauer,  Sebastian and Chumak,  Andrii V.},
  year = {2024},
  month = jul 
}

@article{Cole2023,
	title        = {Negligible magnetic losses at low temperatures in liquid phase epitaxy grown $\mathrm{Y}_3\mathrm{Fe}_5\mathrm{O}_{12}$ films},
	author       = {Will-Cole, A. R. and Hart, James L. and Lauter, Valeria and Grutter, Alexander and Dubs, Carsten and Lindner, Morris and Reimann, Timmy and Valdez, Nichole R. and Pearce, Charles J. and Monson, Todd C. and Cha, Judy J. and Heiman, Don and Sun, Nian X.},
	year         = 2023,
	month        = {05},
	journal      = {Phys. Rev. Mater.},
	publisher    = {American Physical Society},
	volume       = 7,
	pages        = {054411},
	issue        = 5,
	numpages     = 9
}

@article{Kosen2019,
	title        = {Microwave magnon damping in {YIG} films at millikelvin temperatures},
	author       = {Kosen, S. and van Loo, A. F. and Bozhko, D. A. and Mihalceanu, L. and Karenowska, A. D.},
	year         = 2019,
	month        = 10,
	journal      = {APL Mater.},
	volume       = 7,
	number       = 10,
	pages        = 101120,
	doi          = {10.1063/1.5115266}
}

@article{Klingler2017,
	title        = {Gilbert damping of magnetostatic modes in a yttrium iron garnet sphere},
	author       = {Klingler, S. and Maier-Flaig, H. and Dubs, C. and Surzhenko, O. and Gross, R. and Huebl, H. and Goennenwein, S. T. B. and Weiler, M.},
	year         = 2017,
	month        = {03},
	journal      = {Appl. Phys. Lett.},
	volume       = 110,
	number       = 9,
	pages        = {092409},
	doi          = {10.1063/1.4977423},
}

@article{Haidar2015,
	title        = {Thickness- and temperature-dependent magnetodynamic properties of yttrium iron garnet thin films},
	author       = {Haidar, M. and Ranjbar, M. and Balinsky, M. and Dumas, R. K. and Khartsev, S. and \AA{}kerman, J.},
	year         = 2015,
	month        = {03},
	journal      = {J. Appl. Phys.},
	volume       = 117,
	number       = 17,
	pages        = {17D119},
	doi          = {10.1063/1.4914363}
}

@article{Boettcher2022,
	title        = {Fast long-wavelength exchange spin waves in partially compensated {Ga:YIG}},
	author       = {B\"{o}ttcher, T. and Ruhwedel, M. and Levchenko, K. O. and Wang, Q. and Chumak, H. L. and Popov, M. A. and Zavislyak, I. V. and Dubs, C. and Surzhenko, O. and Hillebrands, B. and Chumak, A. V. and Pirro, P.},
	year         = 2022,
	month        = {03},
	journal      = {Appl. Phys. Lett.},
	volume       = 120,
	number       = 10,
	pages        = 102401,
	doi          = {10.1063/5.0082724}
}

@article{Yuan2022,
	title        = {Quantum magnonics: When magnon spintronics meets quantum information science},
	author       = {Yuan, H.Y. and Cao, Yunshan and Kamra, Akashdeep and Duine, Rembert A. and Yan, Peng},
	year         = 2022,
	journal      = {Phys. Rep.},
	volume       = 965,
	pages        = {1–74},
	doi          = {10.1016/j.physrep.2022.03.002},
	type         = {Review},
	publication_stage = {Final},
	source       = {Scopus}
}

@article{Dubs2020,
	title        = {Low damping and microstructural perfection of sub-40nm-thin yttrium iron garnet films grown by liquid phase epitaxy},
	author       = {Dubs, Carsten and Surzhenko, Oleksii and Thomas, Ronny and Osten, Julia and Schneider, Tobias and Lenz, Kilian and Grenzer, J\"org and H\"ubner, Ren\'e and Wendler, Elke},
	year         = 2020,
	month        = {02},
	journal      = {Phys. Rev. Mater.},
	publisher    = {American Physical Society},
	volume       = 4,
	pages        = {024416},
	issue        = 2,
	numpages     = 15
}

@article{Weiler2018,
	title        = {Note: Derivative divide, a method for the analysis of broadband ferromagnetic resonance in the frequency domain},
	author       = {Maier-Flaig, Hannes and Goennenwein, Sebastian T. B. and Ohshima, Ryo and Shiraishi, Masashi and Gross, Rudolf and Huebl, Hans and Weiler, Mathias},
	year         = 2018,
	month        = {07},
	journal      = {Rev. Sci. Instrum.},
	volume       = 89,
	number       = 7,
	pages        = {076101},
	doi          = {10.1063/1.5045135}
}

@article{Chumak2022,
	title        = {Advances in Magnetics Roadmap on Spin-Wave Computing},
	author       = {Chumak, A. V. and Kabos, P. and Wu, M. and Abert, C. and Adelmann, C. and Adeyeye, A. O. and \AA{}kerman, J. and Aliev, F. G. and Anane, A. and Awad, A. and Back, C. H. and Barman, A. and Bauer, G. E. W. and Becherer, M. and Beginin, E. N. and Bittencourt, V. A. S. V. and Blanter, Y. M. and Bortolotti, P. and Boventer, I. and Bozhko, D. A. and Bunyaev, S. A. and Carmiggelt, J. J. and Cheenikundil, R. R. and Ciubotaru, F. and Cotofana, S. and Csaba, G. and Dobrovolskiy, O. V. and Dubs, C. and Elyasi, M. and Fripp, K. G. and Fulara, H. and Golovchanskiy, I. A. and Gonzalez-Ballestero, C. and Graczyk, P. and Grundler, D. and Gruszecki, P. and Gubbiotti, G. and Guslienko, K. and Haldar, A. and Hamdioui, S. and Hertel, R. and Hillebrands, B. and Hioki, T. and Houshang, A. and Hu, C.-M. and Huebl, H. and Huth, M. and Iacocca, E. and Jungfleisch, M. B. and Kakazei, G. N. and Khitun, A. and Khymyn, R. and Kikkawa, T. and Kl\"{a}ui, M. and Klein, O. and K\l{}os, J. W. and Knauer, S. and Koraltan, S. and Kostylev, M. and Krawczyk, M. and Krivorotov, I. N. and Kruglyak, V. V. and Lachance-Quirion, D. and Ladak, S. and Lebrun, R. and Li, Y. and Lindner, M. and Mac\^{e}do, R. and Mayr, S. and Melkov, G. A. and Mieszczak, S. and Nakamura, Y. and Nembach, H. T. and Nikitin, A. A. and Nikitov, S. A. and Novosad, V. and Ot\'{a}lora, J. A. and Otani, Y. and Papp, A. and Pigeau, B. and Pirro, P. and Porod, W. and Porrati, F. and Qin, H. and Rana, B. and Reimann, T. and Riente, F. and Romero-Isart, O. and Ross, A. and Sadovnikov, A. V. and Safin, A. R. and Saitoh, E. and Schmidt, G. and Schultheiss, H. and Schultheiss, K. and Serga, A. A. and Sharma, S. and Shaw, J. M. and Suess, D. and Surzhenko, O. and Szulc, K. and Taniguchi, T. and Urb\'{a}nek, M. and Usami, K. and Ustinov, A. B. and van der Sar, T. and van Dijken, S. and Vasyuchka, V. I. and Verba, R. and Kusminskiy, S. Viola and Wang, Q. and Weides, M. and Weiler, M. and Wintz, S. and Wolski, S. P. and Zhang, X.},
	year         = 2022,
	journal      = {IEEE Trans. Magn.},
	volume       = 58,
	number       = 6,
	pages        = {1--72},
	doi          = {10.1109/TMAG.2022.3149664}
}

@article{SagaYIG,
	title        = {The saga of YIG: Spectra, thermodynamics, interaction and relaxation of magnons in a complex magnet},
	author       = {Cherepanov, V. and Kolokolov, I. and L'vov, V.},
	year         = 1993,
	journal      = {Phys. Rep.},
	volume       = 229,
	number       = 3,
	pages        = {81--144},
	doi          = {10.1016/0370-1573(93)90107-O}
}

@article{Chumak2015,
	title        = {Magnon spintronics},
	author       = {Chumak, A. V. and Vasyuchka, V. I. and Serga, A. A. and Hillebrands, B.},
	year         = 2015,
	journal      = {Nat. Phys.},
	volume       = 11,
	number       = {June},
	pages        = {1505--1549},
	doi          = {10.1007/978-94-007-6892-5\_53},
	isbn         = 9789400768925,
	issn         = {1745-2473}
}

@article{Dubs2017,
	title        = {Sub-micrometer yttrium iron garnet {LPE} films with low ferromagnetic resonance losses},
	author       = {Carsten Dubs and Oleksii Surzhenko and Ralf Linke and Andreas Danilewsky and Uwe Br\"{u}ckner and Jan Dellith},
	year         = 2017,
	month        = {04},
	journal      = {J. Phys. D: Appl. Phys.},
	publisher    = {IOP Publishing},
	volume       = 50,
	number       = 20,
	pages        = 204005,
	doi          = {10.1088/1361-6463/aa6b1c}
}

@article{Lachance-Quirion2019,
	title        = {Hybrid quantum systems based on magnonics},
	author       = {Lachance-Quirion, Dany and Tabuchi, Yutaka and Gloppe, Arnaud and Usami, Koji and Nakamura, Yasunobu},
	year         = 2019,
	journal      = {Appl. Phys. Express},
	publisher    = {IOP Publishing},
	volume       = 12,
	number       = 7,
	pages        = {070101},
	doi          = {10.7567/1882-0786/ab248d}
}

@article{Li2020,
	title        = {Hybrid magnonics: Physics, circuits, and applications for coherent information processing},
	author       = {Li, Yi and Zhang, Wei and Tyberkevych, Vasyl and Kwok, Wai Kwong and Hoffmann, Axel and Novosad, Valentine},
	year         = 2020,
	journal      = {J. Appl. Phys.},
	volume       = 128,
	number       = 13,
	pages        = 130902,
	doi          = {10.1063/5.0020277}
}

@article{Michalceanu2018,
	title        = {Temperature-dependent relaxation of dipole-exchange magnons in yttrium iron garnet films},
	author       = {Mihalceanu, Laura and Vasyuchka, Vitaliy I. and Bozhko, Dmytro A. and Langner, Thomas and Nechiporuk, Alexey Yu. and Romanyuk, Vladyslav F. and Hillebrands, Burkard and Serga, Alexander A.},
	year         = 2018,
	month        = {06},
	journal      = {Phys. Rev. B},
	publisher    = {American Physical Society},
	volume       = 97,
	pages        = 214405,
	issue        = 21,
	numpages     = 9
}

@article{Maier-Flaig2017,
	title        = {Temperature-dependent magnetic damping of yttrium iron garnet spheres},
	author       = {Maier-Flaig, H. and Klingler, S. and Dubs, C. and Surzhenko, O. and Gross, R. and Weiler, M. and Huebl, H. and Goennenwein, S. T. B.},
	year         = 2017,
	month        = {Jun},
	journal      = {Phys. Rev. B},
	publisher    = {American Physical Society},
	volume       = 95,
	pages        = 214423,
	doi          = {10.1103/PhysRevB.95.214423},
	url          = {https://link.aps.org/doi/10.1103/PhysRevB.95.214423},
	issue        = 21,
	numpages     = 8
}

@article{Knauer2023,
	title        = {Propagating spin-wave spectroscopy in a liquid-phase epitaxial nanometer-thick {YIG} film at millikelvin temperatures},
	author       = {Knauer, Sebastian and Dav{\'\i}dkov{\'a}, Krist{\`y}na and Schmoll, David and Serha, Rostyslav O and Voronov, Andrey and Wang, Qi and Verba, Roman and Dobrovolskiy, Oleksandr V and Lindner, Morris and Reimann, Timmy and others},
	year         = 2023,
	journal      = {J. Appl. Phys.},
	publisher    = {AIP Publishing},
	volume       = 133,
	number       = 14,
	pages        = 143905
}

@article{Danilov1989,
	title        = {Low-temperature ferromagnetic resonance in epitaxial garnet films on paramagnetic substrates},
	author       = {Danilov, VV and Lyfar', DL and Lyubon'ko, Yu V and Nechiporuk, A Yu and Ryabchenko, SM},
	year         = 1989,
	journal      = {Sov. Phys. J.},
	publisher    = {Springer},
	volume       = 32,
	pages        = {276--280}
}

@article{Guo2022,
	title        = {Low damping at few-{K} temperatures in $\mathrm{Y}_{3}\mathrm{Fe}_{5}\mathrm{O}_{12}$ epitaxial films isolated from $\mathrm{Gd}_{3}\mathrm{Ga}_{5}\mathrm{O}_{12}$ substrate using a diamagnetic $\mathrm{Y}_{3}\mathrm{Sc}_{2.5}\mathrm{Al}_{2.5}\mathrm{O}_{12}$ spacer},
	author       = {Guo, Side and McCullian, Brendan and Hammel, P Chris and Yang, Fengyuan},
	year         = 2022,
	journal      = {J. Magn. Magn. Mater.},
	publisher    = {Elsevier},
	volume       = 562,
	pages        = 169795
}

@article{Jermain2017,
	title        = {Increased low-temperature damping in yttrium iron garnet thin films},
	author       = {Jermain, C. L. and Aradhya, S. V. and Reynolds, N. D. and Buhrman, R. A. and Brangham, J. T. and Page, M. R. and Hammel, P. C. and Yang, F. Y. and Ralph, D. C.},
	year         = 2017,
	month        = {05},
	journal      = {Phys. Rev. B},
	publisher    = {American Physical Society},
	volume       = 95,
	pages        = 174411,
	doi          = {10.1103/PhysRevB.95.174411},
	issue        = 17,
	numpages     = 5
}

@article{Barak1992,
	title        = {Electron paramagnetic resonance study of gadolinium-gallium-garnet},
	author       = {Barak, J and Huang, MX and Bhagat, SM},
	year         = 1992,
	journal      = {J. Appl. Phys.},
	publisher    = {American Institute of Physics},
	volume       = 71,
	number       = 2,
	pages        = {849--853}
}


\end{document}